\documentclass[ALICE,manyauthors,utphys]{cernphprep}

\usepackage[comma,square,numbers,sort&compress]{natbib}
\usepackage{hyperref}
\usepackage{lineno}

\usepackage{graphicx,type1cm,eso-pic,color}
\usepackage{epsfig}
\usepackage{graphicx}% Include figure files
\usepackage{fancyhdr}
\usepackage{pslatex}   
\usepackage{color}
\usepackage{amsmath, amsthm, amssymb} 
\usepackage{endnotes}
\usepackage{xspace}
\usepackage{multirow}
\usepackage{epstopdf}

\definecolor{grey}{rgb}{0.3,0.3,0.3}
\definecolor{TBD}{rgb}{0.0,0.0,0.5}
\newcommand {\pp} {pp\xspace}

\newcommand {\rootsNN} {\mbox{$\sqrt{s_{\mathrm{NN}}}$}}

\newcommand {\kt} {\mbox{$k_{\mathrm{T}}$}~}

\newcommand {\pt} {\mbox{$p_{\mathrm{T}}$}}
\newcommand {\mt} {\mbox{$m_{\mathrm{T}}$}}

\newcommand {\dnde} { \mbox{$\left< \mathrm{d}N_{\rm ch}/\mathrm{d}\eta \right>^{1/3}$}}
\newcommand {\dndens} { \mbox{$\left< \mathrm{d}N_{\rm ch}/\mathrm{d}\eta \right>^{1/3}$}}

\newcommand {\ktns} {\mbox{$k_{\rm T}$}}

\newcommand {\rout} {\mbox{$R_{\rm out}$}}
\newcommand {\rside} {\mbox{$R_{\rm side}$}}
\newcommand {\rlong} {\mbox{$R_{\rm long}$}}
\newcommand {\dedx} {\mbox{$ \mathrm{d}E/\mathrm{d}x$}\xspace}

\begin{document}%

%%%%%%%%%%%%%%%  Title page %%%%%%%%%%%%%%%%%%%%%%%%
\begin{titlepage}
\PHyear{2015}
\PHnumber{140}      % required, will be obtained from PH
\PHdate{11 June}  % required, will be obtained from PH
%

%%% Put your own title + short title here:
\title{Centrality dependence of pion freeze-out radii \\
in Pb--Pb collisions at $\sqrt{\mathbf{s_{NN}}}$=2.76~TeV}

%\title{Two-pion femtoscopy in p--Pb collisions at $\bf \sqrt{s_{\rm NN}}=5.02$~TeV at the LHC }
\ShortTitle{Centrality dependence of pion freeze-out radii in Pb--Pb collisions}   % appears on right page headers

%%% Do not change the next lines
\Collaboration{ALICE Collaboration\thanks{See Appendix~\ref{app:collab} for the list of collaboration members}}
\ShortAuthor{ALICE Collaboration} % appears on left page headers, do not change

\begin{abstract}
We report on the 
measurement of freeze-out radii for pairs of 
identical-charge pions measured in Pb--Pb collisions at
\rootsNN=2.76~TeV as a function of collision centrality and the
average transverse momentum of the pair \ktns. Three-dimensional sizes
of the system (femtoscopic radii), as well as direction-averaged
one-dimensional radii  are extracted. The radii decrease with  \ktns,
following a power-law behavior. This is qualitatively 
consistent with expectations from a collectively expanding system,
produced in hydrodynamic calculations. The radii also scale
linearly with  \dndens. This behaviour is compared to world data on
femtoscopic radii in heavy-ion collisions. While the dependence is
qualitatively similar to results at smaller \rootsNN, a decrease in
the $R_{\rm out}/R_{\rm  side}$ ratio 
is seen, which is in qualitative
agreement with a specific prediction from hydrodynamic models: a
change from inside-out to outside-in freezeout configuration. The
results provide further evidence for the production of a collective,
strongly coupled system  in heavy-ion collisions at the LHC.
\end{abstract}
\end{titlepage}
\setcounter{page}{2}

\section{Introduction}
\label{sec:intro}

Collisions of lead ions at \rootsNN=2.76~TeV have been recorded
by A Large Ion Collider Experiment (ALICE) at the Large Hadron
Collider (LHC) at CERN. In this energy regime, Quantum Chromodynamics
(QCD) predicts the existence of a new state of strongly interacting
matter, the quark-gluon plasma (QGP) in which quarks and gluons are no
longer confined to individual nucleons. 
Experimental evidence for the existence of such matter has been found
both at the Relativistic Heavy-Ion Collider
(RHIC)~\cite{Adams:2005dq,Adcox:2004mh,Back:2004je,Arsene:2004fa} as
well as at the LHC~\cite{Aamodt:2010cz,Aamodt:2010pa,Abelev:2012di,Aamodt:2010jd,Aad:2010bu,Chatrchyan:2011sx}. %It was found that 
The QGP behaves 
like a fluid with small viscosity and undergoes an explosive
expansion. The study of the space-time structure and dynamics of this process may
reveal important information about the matter properties, such as its
equation of state and the nature of the phase transition between
deconfined and ordinary hadronic matter~\cite{Broniowski:2008vp,Pratt:2009hu}. 

%%%%%%%%%%%%%%%%%%%%%%%%% HBT paragraph

Two-pion correlations at low relative momentum were first shown to be
sensitive to the 
interaction volume
of the emitting source in $\bar{{\rm p}}+{\rm p}$
collisions by G.~Goldhaber et al. 50 years
ago~\cite{Goldhaber:1960sf}. Since then, they were studied in
${\rm e}^{+}+{\rm e}^-$~\cite{Kittel:2001zw}, hadron- and
lepton-hadron~\cite{Alexander:2003ug}, and
heavy-ion~\cite{Adler:2001zd,Adams:2003ra,Adams:2004yc,Adler:2004rq,Lisa:2005dd,Abelev:2006gu,Adler:2006as,Afanasiev:2007kk,Afanasiev:2009ii,Adare:2014vax}
collisions.
Especially in the heavy-ion case, two-particle femtoscopy has 
been developed into a precision tool to probe the
dynamically-generated spatial structure of the emitting system. In
particular, a sharp phase transition between the color-deconfined and
confined states was excluded by the observation of short timescales.
Moreover, femtoscopic measurement, together with other observations
related to bulk collective flow, provide evidence that a
strongly interacting system was created in the
collision~\cite{Hardtke:1999vf,Chojnacki:2007rq,Pratt:2009hu}. 

Femtoscopy in heavy-ion collisions
is understood in
some detail; for example see the experimental overview
in~\cite{Lisa:2005dd} and model calculations
in~\cite{Kisiel:2006is,Chojnacki:2007rq,Soltz:2012rk,Karpenko:2012yf}.
The dependence of the system size extracted from the data is
investigated as  a function of collision centrality and average
transverse momentum of the pair $k_{\rm T} =
|\vec p_{1,\rm T} + \vec p_{2,\rm T}|/2$. As the initial size of the
system grows with increasing   
multiplicity (decreasing centrality), so does the apparent system size
at freeze-out, measured by femtoscopy. Such increase is naturally
produced in a hydrodynamic calculation. Strong hydrodynamic collective
flow in the longitudinal and transverse directions results in the
decrease of the apparent size of the system with increasing \ktns.
This is because longitudinal and transverse velocity boosts cause
particles emitted from spatially separated parts of the collision
region to move away from one another. Such particles cannot have a
small momentum difference, and so correlation functions of boosted
particles are sensitive to only part of the collision region. This
part is referred to as the "homogeneity length"~\cite{Akkelin:1995gh}.
The decrease of the size with \kt is observed 
in experimental data from heavy-ion collisions at all centralities,
various collision energies and colliding system types, and is well
described quantitatively in 
hydrodynamic models~\cite{Broniowski:2008vp,Karpenko:2012yf} and
qualitatively in hadronic rescattering codes~\cite{Li:2012np}. 

Taking into account the successful description of the femtoscopic
scales at lower energies, the hydrodynamic modelling has been
extrapolated to collision energies of the
LHC~\cite{Kisiel:2008ws,Bozek:2011ua,Karpenko:2012yf,Kisiel:2014upa}. The expected
increase 
in initial energy density (temperature) leads to larger evolution
times, which in turn produce larger overall system size and stronger
transverse and longitudinal flows. At the same time the freeze-out
hypersurface 
evolves to have significant positive space-time correlation.
This influences the radii of the system in the plane perpendicular to
the beam axis. In particular the radius along the pair transverse
momentum (called $R_{\rm out}$) is decreased by the correlation with respect
to the other transverse radius (called $R_{\rm side}$), which
decreases the $R_{\rm out}/R_{\rm side}$ ratio
All of those
effects have 
been observed in the first measurement for central (0--5\%)
collisions at the LHC~\cite{Aamodt:2011mr}. This work extends this
measurement to other centralities and compares the obtained radii to
recent hydrodynamic calculations, in order to check their validity in
a large range of event multiplicities. A measurement of
one-dimensional radii was also performed using the two-pion and
three-pion cumulants~\cite{Abelev:2014pja}. This work extends the
two-pion measurement to several ranges of pair transverse momentum. 

The paper is organized as follows. In Section~\ref{sec:datataking} the
data taking conditions and data treatment is described. In
Section~\ref{sec:correlationanalysis} we give the details of the
analysis of the correlation function. In Section~\ref{sec:results} the
extracted radii are presented and compared to model expectations,
while Section~\ref{sec:conclusions} summarizes our findings.

\section{Data taking and track reconstruction}
\label{sec:datataking}

This work reports on the analysis of Pb--Pb collisions produced by
the LHC during the 2010 datataking period. They were recorded by the ALICE
experiment; the detailed description of the detector and the
performance of all of its
subsystems is given in~\cite{Aamodt:2008zz,Abelev:2014ffa}. 
Here we only briefly describe the specific detectors used in this
analysis.
The ALICE Time Projection Chamber (TPC)~\cite{Alme:2010ke} is a large
volume gaseous ionization chamber detector, which was used both for
tracking at mid-rapidity as well as for particle identification via the
measurement of the specific ionization energy loss associated with
each track.  
In addition to the TPC, the information from the ALICE Inner Tracker
System (ITS) was used. The ITS consists of six cylindrical layers, two
Silicon Pixel Detectors closest to the beam pipe, two Silicon Drift
Detector layers in the middle, and two Silicon Strip Detectors on the
outside. The information from ITS was used for tracking and
primary particle selection, as well as for triggering. However the
main triggering detector was the V0. It is a small angle detector
consisting of two arrays of 32 scintillating counters. The first (V0A)
is located 330~cm away from the vertex and covers $2.8<\eta<5.1$, the
second (V0C) is fixed at the front of the hadronic absorber of the
muon arm and covers $-3.7<\eta<-1.7$.
The tracking detectors are
located inside the solenoidal ALICE magnet, which provides a uniform
magnetic field of $0.5$~T along the beam direction. The T0 detector~\cite{Bondila:2005xy}
was a main luminometer in the heavy-ion run. 
It consists of two arrays of 12 Cherenkov counters, covering $-3.28 <
\eta -2.97$ and $4.61 < \eta < 4.92$. It has a time resolution of 40~ps.

The minimum-bias trigger required a signal in both V0 detectors,
which was consistent with the collision occurring at the center of the
ALICE apparatus. A total sample of approximately four million Pb--Pb
events was used for this analysis. For the centrality range considered
in this work the trigger efficiency was 100\%. The centrality was
determined by analyzing the signal from the V0 detector with the 
procedure described in details in~\cite{Aamodt:2010cz}. This
ensured that the centrality determination was obtained using particles
at significantly different rapidities than the ones used for the pion 
correlation analysis. This work presents results 
for seven centrality ranges: 0--5\%, 5--10\%,
10--20\%, 20--30\%, 30--40\%, 40--50\%, and 50--60\% of the 
total hadronic cross section. The position $V_{z}$ of the event vertex
in the beam direction with respect to the center of the ALICE
apparatus was determined for each event. In order to ensure uniform
pseudorapidity acceptance, only events with $|V_{z}|<8$~cm were used
in the analysis. 

For each event, a list of tracks identified as primary pions was
created, separately for positive and negative particles. Each track
was required to leave a signal both in the TPC and the ITS
and the two parts of the track had to match. % to form a single track. 
The TPC is divided by the central electrode into two halves, each of
them is composed of 18 sectors (covering the full azimuthal
angle) with 159 padrows placed radially in each sector. A track signal
in the TPC consists of space points (clusters), each of which is
reconstructed in one of the padrows. 
A track was required to be composed out of at least 80 such
clusters. The parameters of the track are determined by performing a
Kalman fit to a set of clusters. The quality of the fit is judged
by the value of $\chi^{2}$, which was
required to be better than 4 per cluster (each cluster has two
d.o.f.). The transverse momentum of each track was determined from its
curvature in the uniform magnetic field. Two opposite field polarities
were used through the data-taking period, for a check of systematic
tracking effects. The momentum from this fit in the TPC was used in the
analysis. Tracks which had a kink in the trajectory in the TPC were 
rejected. 
Trajectories closer than $3.2$~cm in the longitudinal direction and
$2.4$~cm in the transverse direction to the primary vertex were
selected to reduce the number of secondaries. The kinematic range for
accepted particles was 
$(0.14, 2.0)$~GeV/$c$ in transverse momentum and $(-0.8,0.8)$ in
pseudorapidity. 
Based on the specific ionization energy loss in the TPC gas \dedx, a
probability for each track to be a pion, kaon, 
proton, and electron was determined after comparing with the
corresponding Bethe-Bloch curve. Particles for which the pion
probability was the largest were used in this analysis. This resulted
in an overall purity above 95\%, with small contamination from
electrons in the region where the \dedx curves for
the two particle types intersect. 

The accepted particles from each event are combined into
same-charge pairs. The two-particle detector acceptance
effects must be taken into  account in 
this procedure. Two main effects are present: track splitting and track
merging. Track splitting occurs when a single trajectory is mistakenly
reconstructed as two tracks. The tracking algorithm has been
specifically designed to suppress such cases. In a rare event when
splitting happens, two tracks are reconstructed mostly from the same
clusters in the TPC. Therefore, pairs which share more than 5\%
of clusters are removed from the sample. Together with the
anti-merging cut described below this eliminates the influence of the
split pairs.

Two-particle correlated efficiency and separation power is affected by
track merging. In the TPC, two tracks cannot be
distinguished if their trajectories stay close to each other through
a significant part of the TPC volume. Although this happens rarely,
such pairs by definition have low relative momentum and
therefore their absence distorts the correlation function in the
signal region. The effect of track merging has been studied in central
collisions in the previous work~\cite{Aamodt:2011mr}. In this work we
have used a similar procedure to correct for the merging effects,
through dedicated two-particle selection criteria. More details
are given in Sec.~\ref{sec:systematic}. 

\section{Correlation function analysis}
\label{sec:correlationanalysis}

The two-particle distribution for same-event pion pairs depends on
several factors, including trivial single-particle acceptance
effects. To extract only the relevant two-particle correlation
effects, the correlation function formalism, described below, is
applied. 

\subsection{Correlation function construction}

The femtoscopic correlation function $C$ 
is constructed
experimentally as

\begin{equation}
C(\vec q) = \frac {A(\vec q)} {B(\vec q)},
\label{eq:cfdef}
\end{equation}
where $\vec{q}=\vec{p_{1}}-\vec{p_{2}}$ is the pair relative
momentum (due to fixed masses of the particles only three components
are independent). The magnitude of this vector is referred to as
$q_{\rm inv}$. For a detailed description of the formalism, see
e.g.~\cite{Lednicky:2005af}. 
The signal distribution $A$ is composed of pairs of particles where
both come from the same event. The background distribution $B$ 
is constructed with the ``mixing'' method,
in which the two particles come from different events, which must
have similar characteristics, so that their single-particle efficiency
and distribution are as close as possible. To form a ``mixed'' pair,
particles must come from two events, for which the centralities differ
by no more than 2.5\% and vertex positions differ by no more than
$4$~cm. The correlation function is normalized with the ratio of the
number of pairs in the $B$ and $A$ samples in the full $q$ range used
($0-0.25$~GeV/$c$), so that $C$ at unity means
no correlation. 
The dependence on pair momentum sum is investigated by doing
the analysis for various ranges of \ktns, namely: 
0.2--0.3, 0.3--0.4, 0.4--0.5, 0.5--0.6, 0.6--0.7, 0.7--0.8, and
0.8--1.0~GeV/$c$. The ranges were the same for each centrality range,
which overall resulted in 49 independent correlation functions per
pair charge combination. 

\begin{figure}[htb]
\begin{center}
\vspace*{-.2cm}
\includegraphics*[width=8cm]{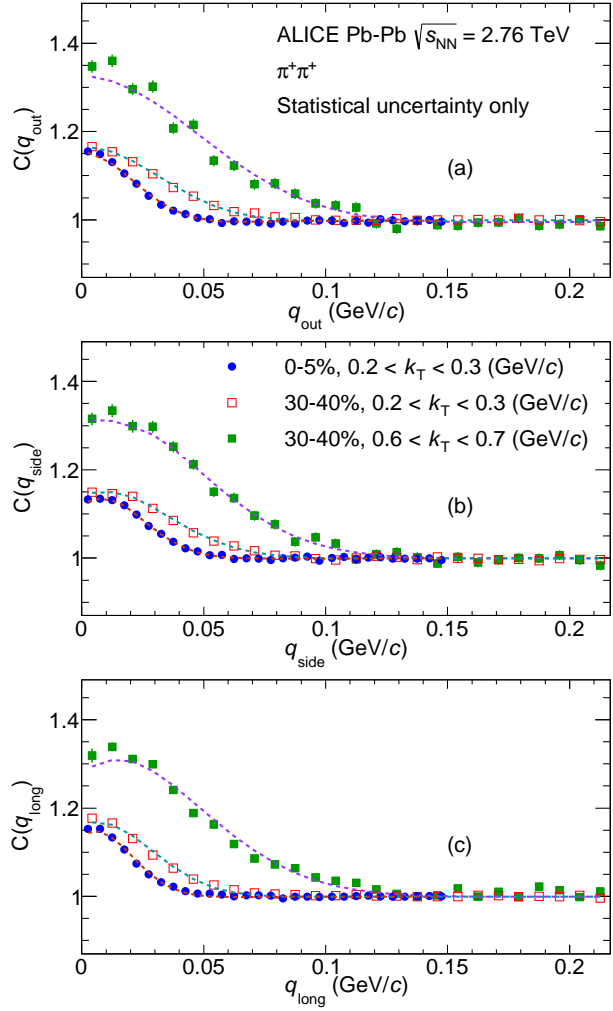} 
\vspace*{-.2cm}
\caption[]{Projections of the Cartesian representation of the
  three-dimensional two-pion correlation 
  functions along the $out$~(a), $side$~(b) and $long$~(c) axes. The
  centrality and pair momentum ranges for the three functions are
  given on the plot. In each case the other components are projected
  over $\pm$ 20 MeV/$c$ around 0 in the other $q$ directions for
  central collisions at low \kt and $\pm$ 33 MeV/$c$ in the other two 
  cases. Lines represent the   corresponding projections of the fit to
  the experimental correlation functions with formula from
  Eq.~\eqref{eq:cfitfun}.} 
\label{fig:fitex3d}
\end{center}
\end{figure}

The momentum difference $\vec{q}$ is calculated in the Longitudinally
Co-Moving System (LCMS), in which the pair total longitudinal momentum
vanishes: $p_{1,L} + p_{2,L}= 0$.  The three coordinates of $\vec{q}$
in LCMS are defined as follows: 
$long$ - along the beam axis, $out$ - along the pair transverse
momentum, and $side$ - perpendicular to the other two. In
Fig.~\ref{fig:fitex3d} the projections of three 
example correlation functions
 along these axes are shown. A
significant, approximately Gaussian enhancement at low 
relative momentum is seen in all projections.
The width of the
correlation grows with increasing centrality (lowering multiplicity)
as well as with increasing $k_{\rm T}$. 

The pair distributions for identical particles have specific
symmetries, which are naturally represented in a spherical harmonic (SH)
decomposition~\cite{Chajecki:2009zg,Kisiel:2009iw}.
In particular for identical
particles all odd-$l$ and odd-$m$ components of a correlation function
representation vanish. The gross features of the correlation function
which are relevant for femtoscopy are fully captured by the low-$l$
components of the decomposition. The $l=0,m=0$ component is sensitive
to the overall size of the pion source. The $l=2,m=0$ component is
sensitive to the difference between longitudinal and transverse size,
and the $l=2,m=2$ component reflects the difference between sideward
and outward component of the transverse radius. Therefore, those three
components of the SH representation contain the same information as
the Cartesian one for
the purpose of the femtoscopic analysis. In particular both
representations allow for fitting of the correlation function with the
same theoretical formula. The next non-vanishing components are for
$l=4$.
Their analysis is beyond the scope of this paper, which focuses
on the overall width (variance) of the distribution in three
directions. 

\begin{figure}[htb]
\begin{center}
\vspace*{-.2cm}
\includegraphics*[width=8cm]{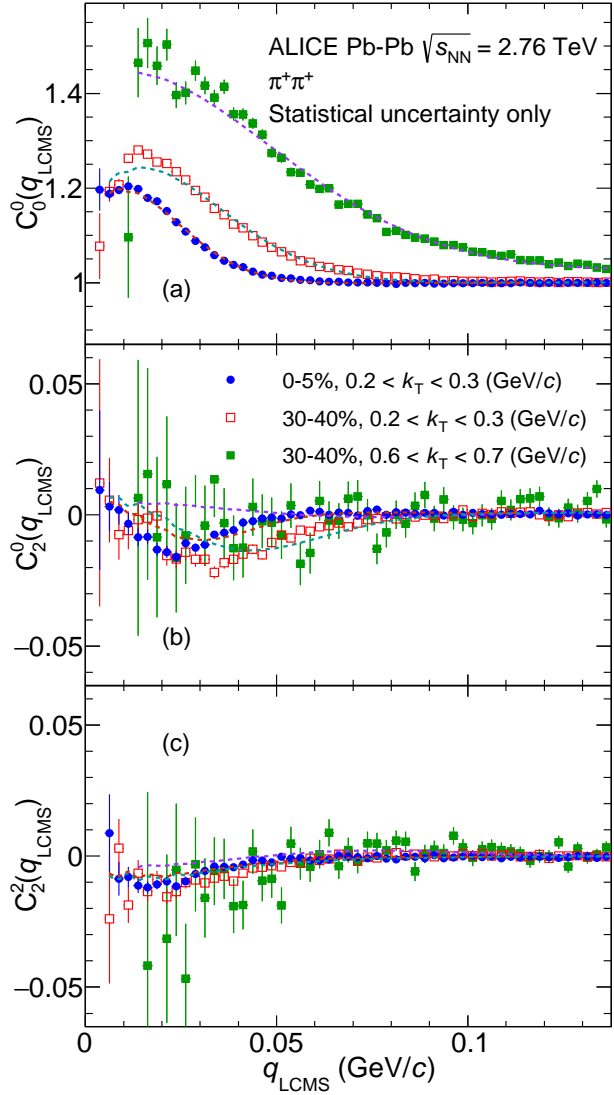} 
\vspace*{-.2cm}
\caption[]{First three non-vanishing components of the SH
  representation of the two-pion correlation functions, $l=0,m=0$~in~(a),
  $l=2,m=0$~in~(b) and $l=2,m=2$~in~(c). The centrality and 
  pair momentum ranges for the three functions are given on the
  plot. The lines show the corresponding SH components of the fit with
  formula from Eq.~\eqref{eq:cfitfun}.}
\label{fig:fitexsh}
\end{center}
\end{figure}

In Fig.~\ref{fig:fitexsh} we show the first three non-vanishing
components of the spherical harmonics representation of three
example correlation functions, the same as in Fig.~\ref{fig:fitex3d}. In the $(0,0)$
component the enhancement at low-$q$ is clearly visible and its width is
increasing with centrality and \ktns. In this representation the low-$q$
dip coming from Coulomb repulsion is visible better than in the
Cartesian one. 
For large \ktns, there is a lack of low-$q$ pairs, as a result of
track merging (see below).
Nevertheless, the correlation clearly extends well beyond the
region with low statistics.
The other two components: $(2,0)$
and $(2,2)$, show a non-trivial correlation structure deviating from
zero, indicating that the shape is not spherical, but rather
ellipsoidal in LCMS. The width of these structures is consistent with
the width of the correlation enhancement  in the $(0,0)$ component,
which means that all three reflect the properties of the femtoscopic
signal. The analysis of the shape of these structures is the main
focus of the next section. 

If the available statistics is limited, as is sometimes the case for
lower collision energies or particles heavier than pions, the analysis
is performed only as a function of magnitude of relative momentum
$q_{\rm inv}$, most naturally calculated in the Pair Rest Frame (PRF). 
In this work we present results in this variable for
completeness. Data was analyzed in the same centrality and pair \kt
ranges as the ones used in the three-dimensional analysis.

\subsection{Fitting the correlation function}
\label{sec:fitting}

The space-time characteristics of the source are reflected in the
correlation function. They are connected via the Koonin-Pratt
equation
\begin{equation}
C(\vec q) = \int S({\bold {r}}, \vec q) |\Psi(\vec q, \bold{r})|^{2}
d^{4}r ,
\label{eq:koonpratt}
\end{equation}
where $\bold{r}$ is the pair space-time separation four-vector. 
$S$ is  the source emission function, interpreted as a
probability to emit a pair of particles with a given relative momentum
and space-time separation. $\Psi$ is the two-particle interaction
kernel. In the simplest case of non-interacting particles
(e.g. photons) it is 
the modulus of the pair wave-function. If the Coulomb or strong
interaction between the particles (called Final State Interaction -
FSI) needs to be taken into account, then $\Psi$ becomes the
Bethe-Salpeter amplitude corresponding to the solution of the relevant
quantum scattering problem, taken with the inverse time 
direction~\cite{Lednicky:2005tb}. 

Previous studies at 
RHIC~\cite{Adler:2001zd,Adams:2003ra,Adams:2004yc,Abelev:2009tp,Adcox:2002uc,Adler:2006as,Afanasiev:2007kk}
and at the LHC~\cite{Aamodt:2011mr} have 
approximated the source by a Gaussian, treating any difference between the
real data and a Gaussian
as a correction. This
procedure was also universally used in all past pion femtoscopic
analyses of heavy-ion collisions. Therefore, we also use it here by
writing 
\begin{equation}
S(\bold r, \vec q) \approx \exp \left(- \frac{r^{2}_{\rm out}}
    {4R^{2}_{\rm out}} - \frac{r^{2}_{\rm side}}
    {4R^{2}_{\rm side}} - \frac{r^{2}_{\rm long}}
    {4R^{2}_{\rm long}} 
 \right )
\label{eq:sfun}
\end{equation}
where $r_{\rm out}$, $r_{\rm side}$, and $r_{\rm long}$ are components
of the relative separation $\mathbf{r}$. 
This static form of $S$ is expressed in LCMS, with $R_{\rm out}$,
$R_{\rm side}$, and $R_{\rm long}$ being the single-particle source
sizes of the system later referred to as ``femtoscopic radii'', or
simply ``radii''. They quantify the lengths of homogeneity of the
system in the $outwards$, $sidewards$, and $longitudinal$ direction,  
respectively.

For like-sign pions the strong interaction contribution is small for
the source sizes expected here (a few~fm)~\cite{Lednicky:2005af}, so
it is neglected. The remaining $\Psi$ is a convolution of the Coulomb
interaction and wave-function symmetrization. As an approximation, 
the Coulomb part is factorized out and integrated separately in the
procedure known as the Bowler-Sinyukov
fitting~\cite{Bowler:1991vx,Sinyukov:1998fc}. It is well tested and is
applicable for pions and for large source sizes 
expected in this analysis. In this approximation the integration of 
Eq.~\eqref{eq:koonpratt} with $S$ given by Eq.~\eqref{eq:sfun} gives the
following fit form for the correlation function

\begin{eqnarray}
C(\vec q) &=& N (1-\lambda) \label{eq:cfitfun} \\ 
&+& N \lambda K_{\mathrm{C}}(q_{\rm inv}) \left [ 1 +
  \exp \left (-R_{\rm out}^{2}q_{\rm out}^{2}-R_{\rm side}^{2}q_{\rm
      side}^{2}-R_{\rm long}^{2}q_{\rm long}^{2} \right)   
\right ] , \nonumber 
\end{eqnarray}
where N is the overall normalization factor. The function $K_{\mathrm{C}}(q_{\rm inv})$
is the Coulomb part of the two-pion wave-function integrated over the
spherical Gaussian source with a given radius. For each correlation
function it is set to the value from the one-dimensional analysis (see
Section~\ref{sec:onedanalysis}), to reflect the 
decrease of the source size with multiplicity and \ktns. Its variation
is a source of systematic uncertainty. The dilution parameter $\lambda$ is
introduced to  account for the fact that not all measured pion pairs
are correlated, and that the real emission function may deviate from a
Gaussian form. 

%\subsection{Performance of the fitting procedure}
The fit is performed with the log-likelihood method for the three
dimensional correlation function in Cartesian representation,
resulting usually in several thousand  degrees of freedom. Examples
are shown in Fig.~\ref{fig:fitex3d}. The Gaussian fit is able to
reproduce the overall width of the correlation in all cases. Some
details of the 
behavior at low-$q$ may not be perfectly described, which can be
attributed to the limitations of the Bowler-Sinyukov formula as well
as to the non-Gaussian, long-range tails which may be present in
the source. Some deviations from the Gaussian ellipsoid shape 
for the higher centrality can 
also be seen for the $long$ direction. We
leave the detailed investigation of these effects for future
work. Nevertheless, the overall sizes $R$ of the system, which are
mostly sensitive to the 
width of the correlation are well estimated. The deviation of the
correlation function from the pure Gaussian shape is %also visibly
smaller than a similar deviation in \pp
collisions~\cite{Aamodt:2011kd}. 

An equivalent fit is also performed for the SH representation of the
correlation. Eq.~\eqref{eq:cfitfun} is numerically integrated
on a $\varphi-\theta$ sphere for each $q_{\rm LCMS}$ bin, with proper
$Y_{l}^{m}$ weights, to produce the 3 components of the SH
decomposition. Statistical uncertainties on each component are taken
into account, as well as the covariance matrix between components.
Examples are shown in Fig.~\ref{fig:fitexsh}. The
fit describes the general direction-averaged width of the correlation 
function, shown in the upper panel. Small deviations can be seen in
the shape and the behavior of the fit at small $q$, as discussed
earlier 
for the Cartesian fit. 
The deviations from zero in the $(2,0)$ and $(2,2)$ components are
small but statistically significant for a large number of
multiplicity/\kt ranges, indicating that the source size is slightly
different in all three directions. 

For the one-dimensional correlation functions constructed as a
function of $q_{\rm inv}$, the fit was performed with a simplified
version of 
Eq.~\eqref{eq:cfitfun}

\begin{eqnarray}
C(\vec q) &=& (1-\lambda) \\ 
&+& \lambda K_{C} \left [ 1 +
  \exp(-R_{\rm inv}^{2}q_{\rm inv}^{2})  
\right ] \nonumber
\label{eq:cfunqinv}
\end{eqnarray}
where the only radius parameter is $R_{\rm inv}$ -- the
one-dimensional direction-averaged femtoscopic radius in the PRF.  

\subsection{Systematic uncertainties of the radii}
\label{sec:systematic}

\begin{table}[tb]
\begin{tabular}{l|c|c|c}
\hline
Uncertainty source & ${\it R}_{\rm out}$ [\%] & ${\it R}_{\rm side}$ [\%] &
${\it R}_{\rm long}$ [\%]  \\
\hline
CF representation & 0.5--5 & 0.5--4 & 0.5--8 \\
Dataset comparison & $<1.5$ & $<1.5$ & $<2$ \\
Fit-range dependence & 0.5--4 & 0.5--3 & 1--5 \\
Two-track cut variation & 3--10 & 2--12 & 2--13 \\
Coulomb correction & 3 & 1 & 1 \\
Momentum resolution correction & 2 & 2 & 2 \\
Centrality estimation & 1 & 1 & 1 \\
\hline
Total & 6--14 & 4--13 & 4--17 \\
\hline
\end{tabular}
\caption{List of contributions to the systematic uncertainty of the
  extracted femtoscopic radii (ranges given). } 
\label{tab:systerror}
\end{table}

Table~\ref{tab:systerror} lists the systematic uncertainty
contributions.
The range of values is given to provide a 
general estimate of the importance of each contribution, however the
systematic uncertainty is estimated for each point 
individually. 
Separate analyses were performed for positive and negative
pions, as well as for two datasets collected with opposite polarities
of the magnetic field inside ALICE. This results in four independent
data samples for which certain systematic effects, most notably the
single-track inefficiencies, are different. Correlation functions for
all four samples for all centrality and pair momentum ranges are
statistically consistent, after all corrections are applied; this is
an important systematic cross-check of the methodology. In the
following discussion the central values are statistical averages of
the fit values obtained for the four samples. The systematic
uncertainty arising from differences among the data sample is between
1 and 2\% for all radii. The other systematic 
uncertainties are analyzed for each sample separately; their final
value is the convolution of the uncertainties for each sample.

Two correlation function representations are used in this work:
the Cartesian and spherical harmonics. They are mathematically
equivalent, the fitting procedure used the same functional form for
both. However the implementation of the fitting procedure is quite
different: log-likelihood vs. regular $\chi^{2}$ fit,
three-dimensional Cartesian histogram vs. three one-dimensional
histograms, fitting range as three-dimensional cube in $q_{\rm
  out},q_{\rm side},q_{\rm long}$ or a three-dimensional sphere with
constant $q_{\rm LCMS}$ radius among others. Therefore, the fits to the
two representations differ systematically upon 
variation of the fitting procedure (fit ranges, Bowler-Sinyukov
approximation, etc.). 
The difference between the values for the two fits is taken as a part
of the systematic uncertainty. It usually ranges from 1\% to 3\% and
grows with pair \kt and multiplicity.

Variation of single particle cuts
around the default value results in modifications of single particle
acceptances and purities. However the correlation function shape
should be to the first order insensitive to those effects. We have
checked that for a reasonable variation of the single particle cuts,
the resulting radii are consistent within statistical uncertainties.

The measurement of the average event multiplicity for a given
centrality range has a known uncertainty of  3--4\% 
for all centrality classes~\cite{Aamodt:2010cz}. The femtoscopic radii
in heavy-ion collsions were observed to scale linearly with \dndens
at lower collision energies~\cite{Lisa:2005dd}. Such trend is also
predicted by hydrodynamical models and is expected to hold at LHC.
Therefore the systematic uncertainty coming from the multiplicity
estimation is about 1\%. 

The dominant systematic effect on the two-particle correlation
function is the two-track correlated efficiency. This effect has been
studied in previous work~\cite{Aamodt:2011mr} for central
collisions. The effect of ``splitting'' 
is fully removed with the help of
dedicated two-track selection criteria, mentioned in
Sec.~\ref{sec:datataking}, and does not influence the radii. However
when two particles' trajectories are located close to each other in
the volume of the TPC detector,
they may be reconstructed as one track or not be reconstructed at
all. This effect, called ``merging'' in the following, results in a
loss of reconstruction efficiency for such pairs of tracks. Pairs of
primary pion trajectories close in space correspond to low relative
momentum, therefore ``merging'' will affect the pion correlation
function exactly in the femtoscopic signal region. The two-track
efficiency was studied 
in Monte-Carlo simulation of the ALICE detector. For pion pairs at low
relative momentum an efficiency loss of up to 20\% was
observed. A two-track selection was chosen, such that the
resulting correlation function was not affected by the
inefficiency. ``Merging'' affects only pairs which are spatially close
in the detector. The ``closeness'' can be quantified by the
pseudorapidity difference $\Delta\eta$ for the pair and only pairs
with $|\Delta\eta|<0.016$ are affected. The trajectories must also be
close in the transverse plane, where they are curved by the magnetic
field. The azimuthal coordinate $\varphi^{*}$ of tracks at a radius of
$1.2$~m from the collision point (so roughly in the center of the TPC
volume) is calculated. Pairs with $\sqrt{\Delta\eta^{2} +
  \Delta{\varphi^{*}}^{2}}<0.045$ are affected by merging. Pairs which
simultaneously satisfy this and the previous angular criteria are
removed both from the signal and the background sample. The correlation
function calculated for pairs surviving this cut is then not affected
by ``merging''. As a systematic check a different procedure to
calculate $\varphi^{*}$ was used, where the value was taken not at a
fixed radius, but instead at a radius inside the TPC (i.e. between $0.5$
and $2.4$~m) where $\varphi^{*}$ was the smallest. The differences
between radii for the correlation functions calculated with the two
methods is taken as systematic uncertainty. It is from 2--13\%. 

A pair of same charge pions traversing a solenoidal field can have
two configurations: a ``sailor'' where the two trajectories bend away
from one another and never cross, or a ``cowboy'' where the trajectories
bend toward each other and can cross~\cite{Adams:2004yc}. Merging affects the
``cowboy'' pairs, but only weakly influences the ``sailors''. The two
configurations correspond to different phase space regions in the
Cartesian representation of the correlation function. We have
performed two fits to the correlations (corrected for merging),
restricting the fitting range to either the ``cowboy'' or the ``sailor''
region. The consistency of the radii obtained in these fits was used
as an estimator of the effectiveness of the cut for removing merged
tracks. A less restrictive cut degraded agreement between the ``cowboy''
and ``sailor'' radii, while making the cut stricter did not improve the
agreement but reduced statistics in the signal region. The comparison
allowed optimization of the cut, and provided another estimate of the
systematic uncertainty on the two-track correction procedures. The
uncertainty estimated in this way is consistent with the uncertainty
determined by varying the $\varphi^{*}$ definition. This uncertainty is largest
for transverse radii, is most prominent for high multiplicity events
(central collisions), and affects a wider region in $q$ for pairs with
higher \ktns. 

The separation into the ``cowboy'' and ``sailor'' phase-space regions
is not feasible for the SH representation of the correlation
function. In this case, when the anti-merging cut is not properly 
applied, one observes significant non-femtoscopic signals, especially
in the $C_{2}^{2}$ component of the correlation function and a
reliable fit cannot be performed. Therefore, the $C_{2}^{2}$ component
serves as a sensitive independent check of the effectiveness of the
``anti-merging'' cut. 
This is a good illustration on
how the two representations complement each other in the systematic
studies. 

The fit was performed for several values of the fitting range in $q$
(varying with multiplicity and pair \ktns, following the changes in
the correlation function width). The variation of the fitted radii
with the change of the range was taken as another component of the
uncertainty, which is less than 5\% for all the radii.   

In addition to the uncertainties listed above, other systematic
effects can influence the extracted radii. The first is the momentum
resolution, which was studied in~\cite{Aamodt:2011mr}. The correction
procedure described there is used in this work, as well. The
uncertainty on the radii from this correction is 2\%. Another effect
is the influence of the Bowler-Sinyukov procedure on the extracted
radii and $\lambda$ parameter (fraction of correlated pairs). The
procedure results in an uncertainty of 3\% on \rout, 1\% on the other
radii, and lowers the $\lambda$ value by up to
5\%~\cite{Kisiel:2006is}.  

All the systematic uncertainty components mentioned above are added in
quadrature, the range of values of the total systematic uncertainty is
given in Table~\ref{tab:systerror}. 

The one-dimensional analysis is performed in
PRF, where the total momentum of the pair vanishes. In the
transformation from LCMS to PRF $R_{\rm out}$ is scaled by the
$\gamma$ factor for the pair, depending on \kt and as a consequence is
larger than the other two components. In such case the
direction-averaged one-dimensional correlation function  
becomes non-Gaussian. This produces a dependence of the fit value on
the range of the fit, resulting in a systematic uncertainty of
up to 10\%. Other components of the uncertainty, such as field
orientation dependence, momentum resolution and Coulomb correction
dependence are comparable to those from the three-dimensional fit.

\section{Results}
\label{sec:results}
\subsection {Three-dimensional radii}

\begin{figure}[htb]
\begin{center}
\vspace*{-.2cm}
\includegraphics*[width=8cm]{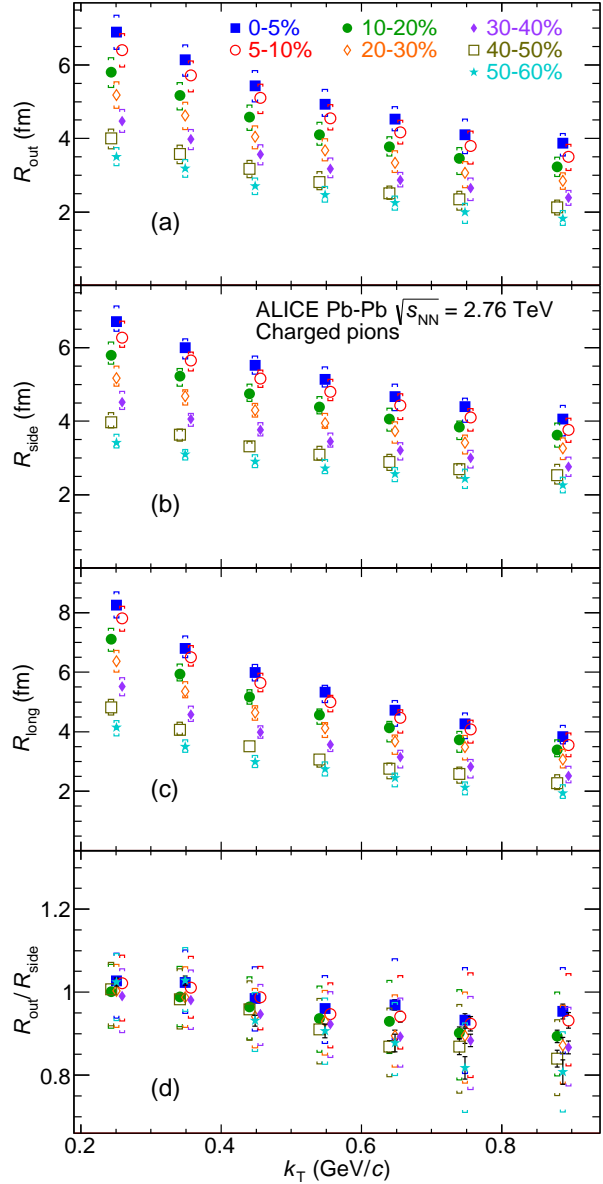} 
\vspace*{-.2cm} 
\caption[]{Femtoscopic radii as a function of pair transverse momentum
  \kt for seven centrality ranges. $R_{\rm out}$ is shown in (a),
  $R_{\rm side}$ in (b), $R_{\rm long}$ in (c) and $R_{\rm out}/R_{\rm
    side}$ in (d). The points for centralities 5--10\%, 10--20\%,
  30--40\% and 40--50\% have been slightly shifted in the x direction
  for visibility. Statistical uncertainties are shown as black lines
  (usually smaller than symbol size), systematic uncertainties are shown
  as colored caps.} 
\label{fig:pbktcentradii4p}
\end{center}
\end{figure} 

The outcome of the fitting procedure are 49 sets of femtoscopic radii,
one set for each centrality and \kt range. They are shown in panels
a)--c) of Fig.~\ref{fig:pbktcentradii4p}. The radii in all directions
are in the range of 2 to 8.5 fm. The radii universally decrease with
increasing 
\ktns, in qualitative agreement with a decreasing homogeneity length,
as predicted by hydrodynamics.
Such
behavior is a strong indication of a large degree of collectivity in
the created system. The radii are also universally higher for more
central collisions,
which correspond to growing final-state
event multiplicity. For the lowest \ktns, $R_{\rm long}$ is generally the
largest, whereas at large \kt there is no universal ordering of the
radii.  

In the most central collisions the values of the $\lambda$ parameters are
around 0.33 for the lowest \kt range, are increasing linearly to 0.43 for \kt of
$0.65$~GeV/$c$, and falling to 0.36 in the last \kt range. 
For peripheral collisions, they are higher by 0.05 to 0.07. 
The lowering of $\lambda$ from a theoretical maximum of 1.0 can be
attributed to a number of factors. 
The pion sample contains daughters of long-lived strongly decaying
resonances; their fraction falls with growing \pt. As a result the
source function contains non-Gaussian tails extending to large
relative separation. There may be additional factors influencing the
shape of the correlation function, coming from the source
dynamics. The exact source shape usually deviates from a Gaussian, in
a way that lowers the $\lambda$ parameter of the Gaussian fit
significantly. The detailed study of this shape requires a dedicated
methodology and is beyond the scope of this work. 
In addition, at the largest \kt the electron and pion \dedx become
comparable in the TPC, and the pion sample is contaminated by
electrons, which also lowers $\lambda$. The approximate
treatment of the Coulomb interaction in the Bowler-Sinyukov fitting
procedure lowers it by 5--10\%~\cite{Kisiel:2006is}. This effects is
most pronounced for large source sizes (central collisions).  
Finally, possible coherent emission of pions~\cite{Abelev:2013pqa}
is expected to lower $\lambda$ by a few percent.  
The $\lambda$ parameter values given here are for
reference only and their physics interpretation is not discussed. 

In panel (d) of Fig.~\ref{fig:pbktcentradii4p} the $R_{\rm out}/R_{\rm
  side}$ ratio is shown. Its systematic uncertainty is determined
independently from those of $R_{\rm out}$ and $R_{\rm side}$, to
account for the fact that they may be correlated. The ratio is
consistent with unity for central collisions.
Its value slowly decreases for more peripheral collisions, and reaches
$~$0.85 for peripheral collisions and high \ktns. 
Based on hydrodynamic models, the $R_{\rm out}/R_{\rm side}$ ratio has
been proposed as a sensitive probe of the shape and space-time
correlation present at the freeze-out
hypersurface~\cite{Kisiel:2008ws,Karpenko:2009wf}. In particular this
ratio at the LHC was predicted to be lower than a value of $1.1$
measured at top RHIC collision energies.

\begin{figure}[htb] 
\begin{center}
\vspace*{-.2cm}
\includegraphics*[width=8cm]{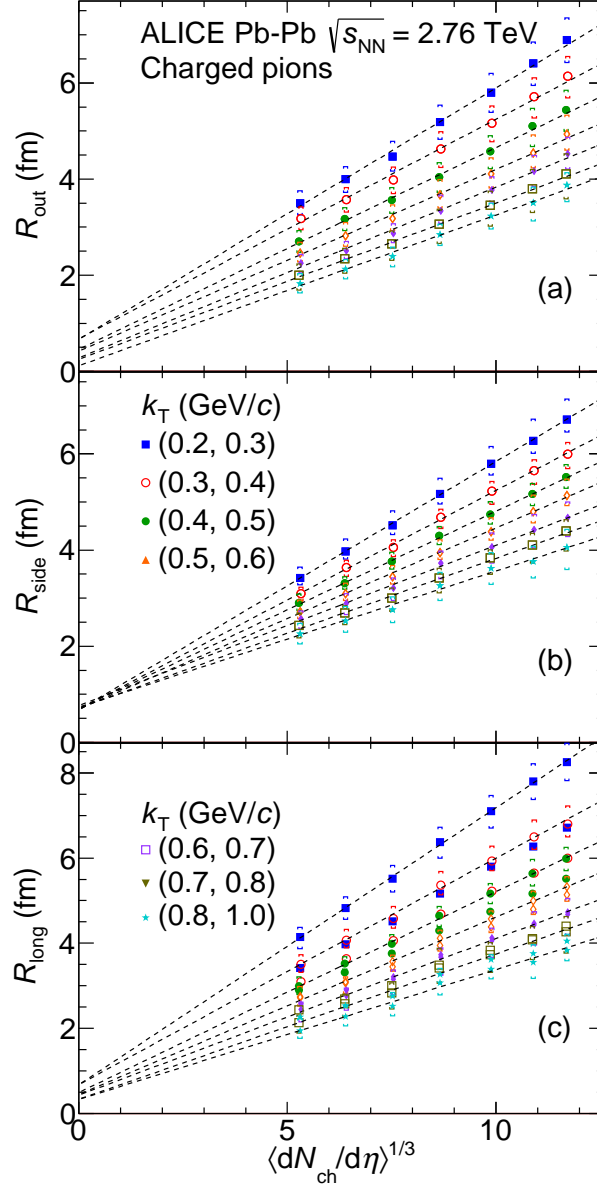} 
\vspace*{-.2cm}
\caption[]{The radii plotted as a function of \dnde ~for seven \kt
  ranges. $R_{\rm out}$ is plotted in 
(a), $R_{\rm side}$ in (b), and $R_{\rm long}$ in  (c). The
dashed lines represent linear fits to each set of points for a given
component and \kt range. Points for some \kt ranges have been 
slightly shifted in the $x$ direction for visibility. Statistical
uncertainties are shown as black lines (usually smaller than symbol
size), systematic uncertainties are shown as colored caps.}
\label{fig:pbradmult}
\end{center}
\end{figure} 

\subsection{Scaling of the radii}

It has been argued in~\cite{Lisa:2005dd} that the femtoscopic volume
scales with the final-state event multiplicity, and that each of the
three-dimensional radii separately scales with this value taken to the
power $1/3$. In Fig.~\ref{fig:pbradmult} we present the dependence of
the radii on multiplicity
for Pb--Pb collisions. The scaling is evident for all
datasets, for all three directions and all analyzed pair momentum
ranges.  

\begin{figure}[htb]
\begin{center}
\vspace*{-.2cm}
\includegraphics*[width=8cm]{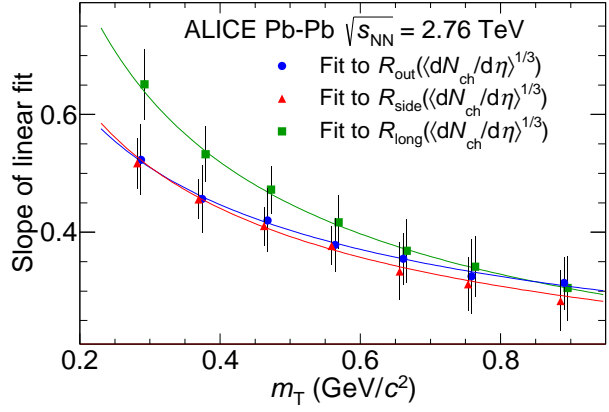} 
\vspace*{-.2cm}
\caption[]{The slope parameters of the linear fits shown in
  Fig.~\ref{fig:pbradmult}, as a function of pair \mt. Lines represent
  the power-law fits (see text for details). Errorbars represent the
  uncertainty of the parameter of the fits, in which the combined
  systematic and statistical uncertainties on the radii ware taken
  into account.} 
\label{fig:pbradamt}
\end{center}
\end{figure} 

Similarily, hydrodynamics predicts approximate scaling of the radii
with pair transverse mass $m_{\rm T}=\sqrt{k_{\rm T}^{2} +
  m_{\pi}^{2}}$~\cite{Kisiel:2014upa}. The slope parameters of the
lines shown in Fig.~\ref{fig:pbradmult} are plotted in
Fig.~\ref{fig:pbradamt}, as a function of \mt.
They
are fitted with a power law function of the form

\begin{equation}
a(m_{\rm T}) = \beta \left( \frac{m_{\rm T}} {m_{\pi}}
\right)^{\alpha}
\label{eq:mtplfit}
\end{equation}
where $\beta$ and $\alpha$ are free parameters. The slope parameters
follow the power law scaling within the current systematic
uncertainties, the value of the $\alpha$ parameter is $-0.65 \pm 0.12$
for $long$, $-0.46 \pm 0.13$ for $out$, and $-0.52 \pm 0.11$ for $side$
direction. The dependence of the values of femtoscopic
radii on centrality and \kt factorizes into a linear dependence on
\dnde~ and a power law dependence 
on \mt.

\subsection{One-dimensional analysis}
\label{sec:onedanalysis}

\begin{figure}[htb]
\begin{center}
\vspace*{-.2cm}
\includegraphics*[width=8cm]{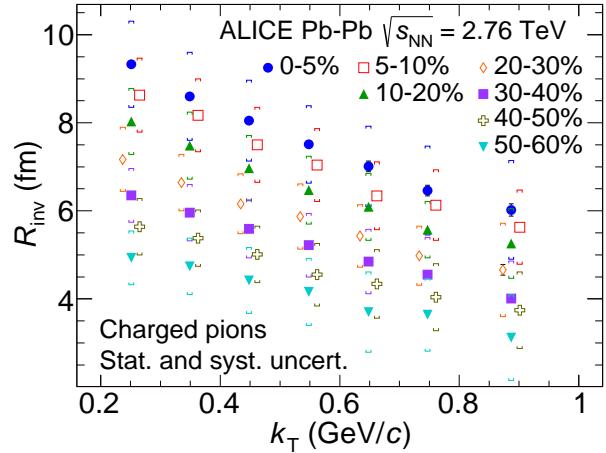} 
\vspace*{-.2cm}
\caption[]{One-dimensional femtoscopic radius $R_{\rm inv}$ as a function
  of pair transverse momentum \kt for seven centrality ranges. Points
  for 5--10\%, 20--30\%, and 40--50\% centrality have been slightly
  shifted in the $x$ direction for visibility. Statistical uncertainty
  is shown as black lines, systematic uncertainty is shown as colored caps.}
\label{fig:1dradii}
\end{center}
\end{figure} 

\begin{figure*}[htb]
\begin{center}
\vspace*{-.2cm}
\includegraphics*[width=15cm]{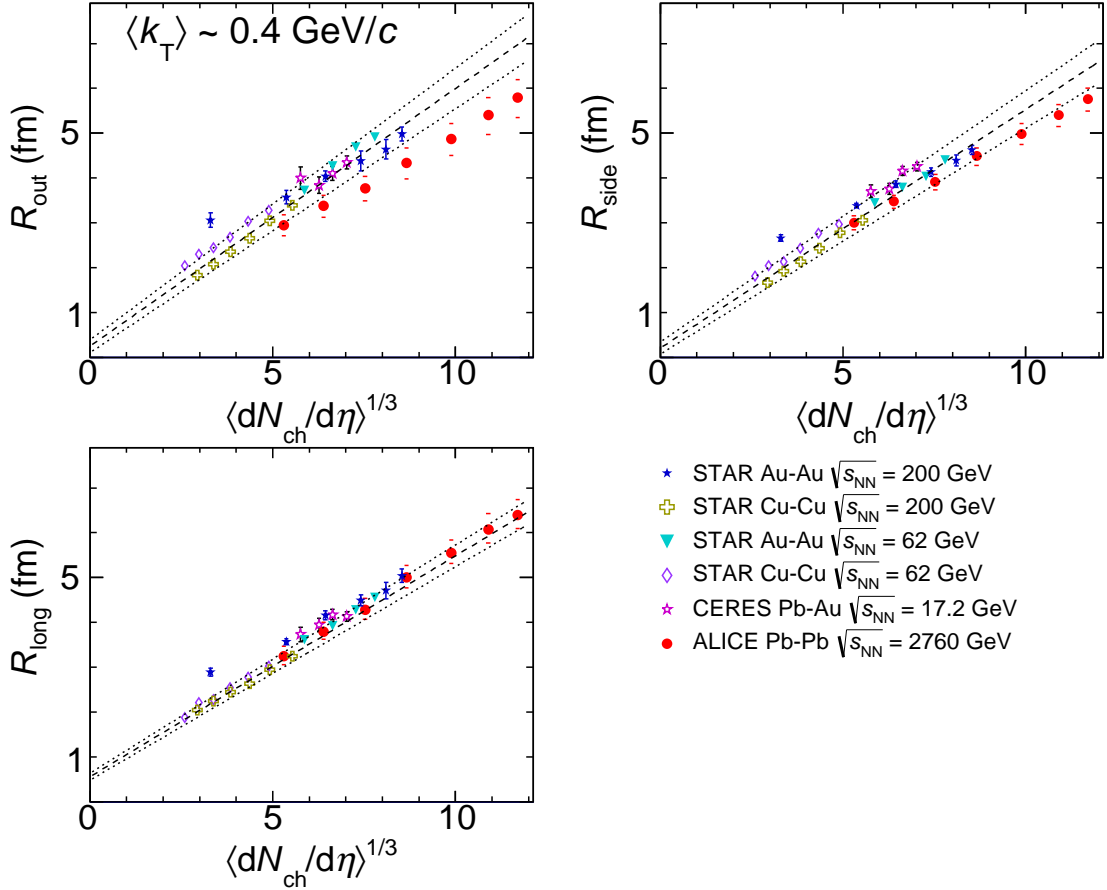} 
\vspace*{-.2cm} 
\caption[]{Comparison of femtoscopic radii, as a function of measured
  charged particle multiplicity, for a number of collision systems and
  collision
  energies~\cite{Adamova:2002wi,Adams:2004yc,Abelev:2009tp}. 
  Dashed lines show linear fits done to heavy-ion data, excluding
  ALICE (dotted lines show one sigma contour). Various experiments use
  differing \kt ranges, values shown on the plot are for the range
  for which the average \kt is closest to the selected value of
  $0.4$~GeV/$c$; in case of ALICE an average of two neighbouring
  ranges of \kt is shown. Systematic uncertainties are shown where
  available.
}
\label{fig:pbtopcomp}
\end{center}
\end{figure*} 

The results of the one-dimensional fits with Eq.~\eqref{eq:cfunqinv}
are shown in Fig.~\ref{fig:1dradii}. Similarly to the
three-dimensional case, the radius is increasing with event
multiplicity (decreasing centrality). Therefore, the final state shape
is reflecting the growth of the initial shape with decreasing
centrality. The $R_{\rm inv}$ is also decreasing with pair transverse
momentum. This is usually understood as a manifestation of the hydrodynamic
collectivity. The one-dimensional radius also serves 
as a comparison basis with the femtoscopic analysis for heavier
particles, where the one-dimensional analysis is standard and the
three-dimensional analysis is challenged by the more complicated
description of the pair interaction as well as significantly smaller
statistics~\cite{Abelev:2012ms,Abelev:2012sq}.

\subsection{Comparison to previous measurements}

In Fig.~\ref{fig:pbtopcomp} the heavy-ion data from Pb--Pb collisions
at the LHC reported in this work are compared to the previous
measurements, including results obtained at lower collision energies. 
It has been argued~\cite{Lisa:2005dd} that three-dimensional
femtoscopic radii scale with the cube root of measured  charged particle
multiplicity not only for a single energy and collision system, but
universally, across all collision energies and initial system
sizes. 
The dashed lines in the figure are linear fits to heavy-ion
data available before the startup of the LHC (the dotted lines show
one sigma contours of these fits). 
At lower energies the linear scaling
was followed well in $long$ and $side$ directions and only
approximately in $out$, with some outliers such as the most peripheral
collisions reported by STAR. Our data at higher
collision energy show that the scaling in the $long$ direction 
is preserved. The data for the $side$ direction fall below the scaling 
trend, although still within the statistical uncertainty. A clear
departure from the linear scaling is seen in the $out$ direction; data
from the LHC lie clearly below the trend from lower energies. Such behavior
was predicted by hydrodynamic calculations~\cite{Kisiel:2008ws} and is
the result of the modification of the freeze-out shape. Larger initial
deposited 
energy produces larger temperature gradients and longer evolution time
at LHC. This results in a change from outside-in to inside-out  
freeze-out and this modification of the space-time correlation drives
the $R_{\rm out}/R_{\rm side}$ ratio to values lower than at RHIC.
Therefore, already for
heavy-ion data the simple universal linear scaling is broken in the
transverse direction. As already reported in~\cite{Aamodt:2011kd}, the
femtoscopic radii for \pp collisions also exhibit linear scaling in
the same variable, albeit with significantly different parameters. In
this case the scaling between different colliding systems is broken
again in the longitudinal direction. 

Other scaling variables were proposed for the femtoscopic
radii, based on Monte-Carlo simulations of the initial state, such as
average number of participants $N_{\rm part}$~\cite{Abelev:2009tp} or
the characteristic initial transverse size
$\bar{R}$~\cite{Adare:2014vri,Adare:2014pta}. If data from different
centralities but the same collision energy are considered, a linear
scaling is indeed observed for radii in all directions, if they are
plotted as a function of either of these variables. However, 
no such scaling is observed when
data from two collision energies (e.g. top RHIC and top LHC) is
compared. In that sense these variables are less adequate than \dndens,
for which the linear scaling is preserved, across many collision
energies, for at least one direction ($R_{\rm long}$). This
observation is consistent with an expectation that the final
freeze-out volume, reflected in the femtoscopic radii, should scale
with the final-state observable (such as e.g. \dndens), while the simple
``geometric'' initial state variables do not contain enough
information. They must be replaced with modeling of the full
evolution process, such as the one given by hydrodynamics, which
depends on additional parameters apart from the initial size, such as
initial energy density.  

\subsection{Model comparisons}

The hydrodynamic models predict both the centrality and pair momentum
dependence of the femtoscopic radii. Usually the parameters of the
model (initial energy density profile, the equation of state and the
freeze-out condition) are adjusted to reproduce the shape of the
single particle inclusive transverse momentum spectra as well as the
elliptic flow. The charged particle multiplicity must also be
reproduced, preferably as a function of pseudorapidity if the model
employs full three-dimensional modeling. The total particle
multiplicity is usually determined on the freeze-out hypersurface by
employing statistical hadronization. After converting the continuous
medium to hadrons, final state interactions are taken into account
either in the simplified form, with propagation and decay of hadronic
resonances or with the full rescattering simulation. We compare our
results to predictions from the Therminator model coupled to (3+1)D
viscous hydrodynamics~\cite{Bozek:2011ua,Bozek:2012qs,Kisiel:2014upa}.
Similar results have been obtained in the HKM
model~\cite{Karpenko:2012yf}. 

\begin{figure}[htb]
\begin{center}
\vspace*{-.2cm}
\includegraphics*[width=8cm]{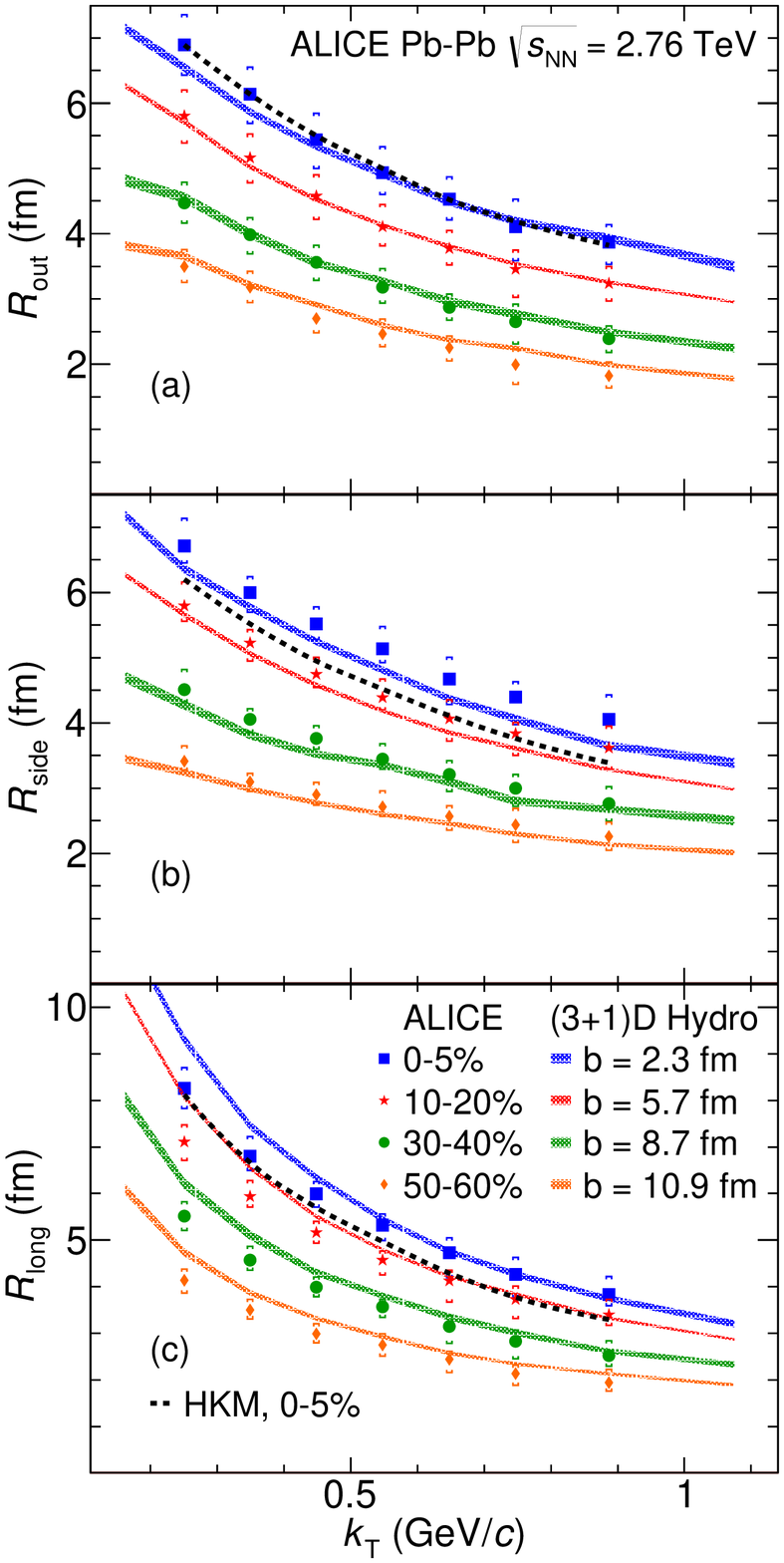} 
\vspace*{-.2cm}
\caption[]{Comparison of the femtoscopic radii (\rout ~in (a), \rside~
  in (b), and \rlong~ in (c)), as a function of pair
  transverse momentum, with the calculation from the Therminator and
  (3+1)D hydro model~\cite{Kisiel:2014upa}, for four
  centralities (here identified by the value of the impact parameter
  $b$ used in the calculation) and with the HKM
  model~\cite{Karpenko:2012yf} for the central data. Closed symbols
  are experimental data (with statistical and systematic uncertainty),
  bands and dashed lines are the model 
  calculations.}  
\label{fig:pbtoth3vcomp}
\end{center}
\end{figure} 

\begin{figure}[htb]
\begin{center}
\vspace*{-.2cm}
\includegraphics*[width=8cm]{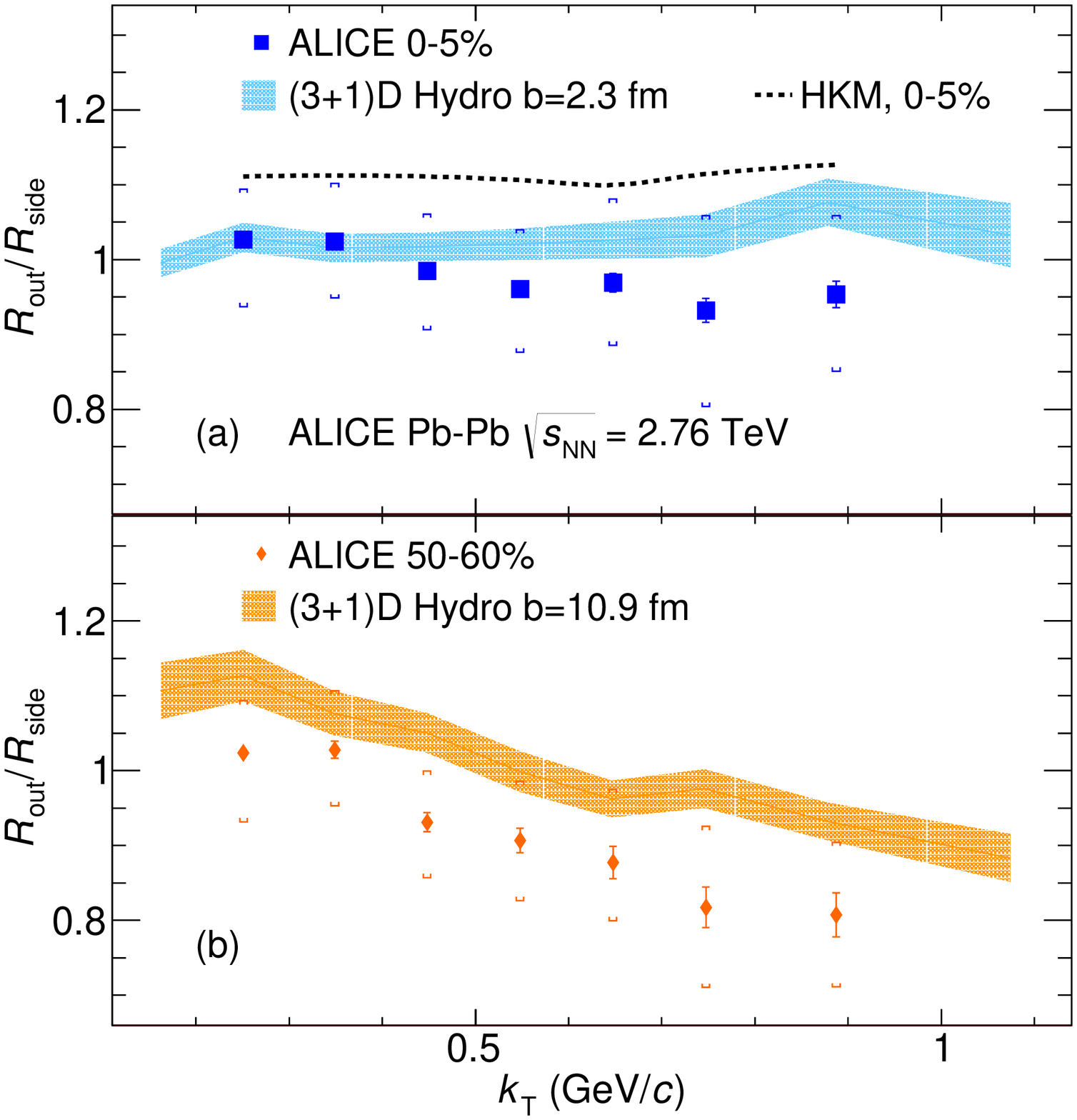} 
\vspace*{-.2cm}
\caption[]{Comparison of the $R_{\rm out}/R_{\rm side}$ ratio, as a
  function of pair transverse momentum, with the calculation from the
  Therminator and (3+1)D hydro model~\cite{Kisiel:2014upa}, for 0--5\%
  centrality in panel (a) and 50--60\% centrality in panel (b). The
  comparison to the HKM model~\cite{Karpenko:2012yf} is also shown for
  the central data. Closed symbols are experimental data (with
  statistical and systematic uncertainty), bands and
  dashed lines are the model calculations.}  
\label{fig:pbtothoscomp}
\end{center}
\end{figure} 

In Fig.~\ref{fig:pbtoth3vcomp} we show the comparison of our data to
the calculations from the (3+1)D hydrodynamic model coupled to the
Therminator statistical hadronization code. The model is fully
three-dimensional and it is able to reproduce values of $R_{\rm long}$ 
for all centralities, with some overprediction of the overall
magnitude and the slope of the \kt dependence, especially at low
momentum. This is an indication that the longitudinal dynamics is
reasonably described in the model, both in momentum and space-time
sectors. 
The $R_{\rm out}$ is well described for all centralities,
both in magnitude as well in the slope of the \kt dependence. 
The slope of the \kt dependence is also well described for $R_{\rm
  side}$, but the magnitude is lower than in data, although within
the systematic uncertainty. 
The 
intercept of $R_{\rm side}$ at low \kt is usually associated with the
overall geometrical size of the system, while the slope of the \kt
dependence of both transverse radii depends on the amount of flow in
the system. 
Both are well reproduced, so the hydrodynamic approach is in good
agreement with our data. 

Another model based on hydrodynamic formalism, the
HKM~\cite{Karpenko:2012yf} is also shown in
Fig.~\ref{fig:pbtoth3vcomp} for 0--5\% most central collisions. It
differs from  the previous in the implementation of the freeze-out
process. It also directly treats hadron rescattering with the UrQMD
simulation. Nevertheless, the pion femtoscopy in central Pb--Pb
collisions at the LHC is reproduced in the calculation, with some
underprediction of $R_{side}$ and $R_{long}$ magnitude. Therefore, the
approximate agreement with data is 
a universal feature of such models, not of a particular implementation. 
The particular choice of initial conditions and the equation of state
for these models was motivated by the analysis of RHIC femtoscopic
data. 
This choice is essential for the correct description of the data,
due to the collective flow mechanism in
the hydrodynamic evolution at the LHC energies.
It also indicates that the details of the freeze-out process have
limited influence on pion femtoscopy. Some studies
suggest that femtoscopy of heavier particles
might be a more sensitive probe in this case~\cite{Karpenko:2012yf}.

The data to model comparison of the $R_{\rm
  out}/R_{\rm side}$ ratio is plotted in
Fig.~\ref{fig:pbtothoscomp}. It  shows values consistent with unity
for central collisions, both for 
models and data. Such low values are associated with the change to
outside-in freeze-out scenario at LHC collision energies.  For most
peripheral collisions the ratio decreases with \kt even more, to
values smaller than unity. This decrease is qualitatively reproduced
in the model, although the calculations are at the upper edge of the 
experimental systematic uncertainty.

\section{Conclusions}
\label{sec:conclusions}

We reported on the  centrality and pair \kt
dependence of the three-dimensional and direction-averaged
one-dimensional pion femtoscopic radii in Pb--Pb collisions at 
\rootsNN=2.76~TeV.  The behavior of the femtoscopic radii can be
factorized into a linear dependence on 
cube root of charged particle density (separately for each of the
three-dimensional radii) and a power-law dependence on pair transverse
momentum, with slightly different exponents for each direction. The
dependence for the longitudinal radius is steeper than for the transverse
ones. The dependence was also compared to the one observed in heavy-ion
collisions at lower energies and to other collisions systems. The
 radii at the LHC follow the ``universal'' \dnde~ scaling in the $long$
direction, but the radii in the transverse directions are below the
``universal'' curve. Simple linear scaling predictions are not valid
when the collision energy is increased by an order of magnitude. The
details of the dynamic evolution of the system influence the results
significantly. This is in qualitative agreement with predictions from 
hydrodynamic models. In particular, when moving from RHIC to LHC
collision energies, they produce a change in freeze-out shape, larger
transverse radial flow and longer system evolution time. 
Comparison of the full dataset to the calculations from the recent
hydrodynamic models, including three-dimensional evolution as well as
hadronic stage, generally show a good agreement, which is complementary
to similar agreement observed for momentum-only observables, such as
momentum spectra and elliptic flow. The existence of such agreement
both in the space-time as well as in momentum sectors, provides strong
arguments for the validity of hydrodynamic models for the description of
flowing bulk matter created in heavy-ion collisions at the LHC. 

%%%%% acknowledgements
\newenvironment{acknowledgement}{\relax}{\relax}
\begin{acknowledgement}
\section*{Acknowledgements}
The ALICE Collaboration would like to thank all its engineers and technicians for their invaluable contributions to the construction of the experiment and the CERN accelerator teams for the outstanding performance of the LHC complex.
The ALICE Collaboration gratefully acknowledges the resources and support provided by all Grid centres and the Worldwide LHC Computing Grid (WLCG) collaboration.
The ALICE Collaboration acknowledges the following funding agencies for their support in building and
running the ALICE detector:
State Committee of Science,  World Federation of Scientists (WFS)
and Swiss Fonds Kidagan, Armenia,
Conselho Nacional de Desenvolvimento Cient\'{\i}fico e Tecnol\'{o}gico (CNPq), Financiadora de Estudos e Projetos (FINEP),
Funda\c{c}\~{a}o de Amparo \`{a} Pesquisa do Estado de S\~{a}o Paulo (FAPESP);
National Natural Science Foundation of China (NSFC), the Chinese Ministry of Education (CMOE)
and the Ministry of Science and Technology of China (MSTC);
Ministry of Education and Youth of the Czech Republic;
Danish Natural Science Research Council, the Carlsberg Foundation and the Danish National Research Foundation;
The European Research Council under the European Community's Seventh Framework Programme;
Helsinki Institute of Physics and the Academy of Finland;
French CNRS-IN2P3, the `Region Pays de Loire', `Region Alsace', `Region Auvergne' and CEA, France;
German Bundesministerium fur Bildung, Wissenschaft, Forschung und Technologie (BMBF) and the Helmholtz Association;
General Secretariat for Research and Technology, Ministry of
Development, Greece;
Hungarian Orszagos Tudomanyos Kutatasi Alappgrammok (OTKA) and National Office for Research and Technology (NKTH);
Department of Atomic Energy and Department of Science and Technology of the Government of India;
Istituto Nazionale di Fisica Nucleare (INFN) and Centro Fermi -
Museo Storico della Fisica e Centro Studi e Ricerche "Enrico
Fermi", Italy;
MEXT Grant-in-Aid for Specially Promoted Research, Ja\-pan;
Joint Institute for Nuclear Research, Dubna;
National Research Foundation of Korea (NRF);
Consejo Nacional de Cienca y Tecnologia (CONACYT), Direccion General de Asuntos del Personal Academico(DGAPA), M\'{e}xico, :Amerique Latine Formation academique – European Commission(ALFA-EC) and the EPLANET Program
(European Particle Physics Latin American Network)
Stichting voor Fundamenteel Onderzoek der Materie (FOM) and the Nederlandse Organisatie voor Wetenschappelijk Onderzoek (NWO), Netherlands;
Research Council of Norway (NFR);
National Science Centre, Poland;
Ministry of National Education/Institute for Atomic Physics and Consiliul Naţional al Cercetării Ştiinţifice - Executive Agency for Higher Education Research Development and Innovation Funding (CNCS-UEFISCDI) - Romania;
Ministry of Education and Science of Russian Federation, Russian
Academy of Sciences, Russian Federal Agency of Atomic Energy,
Russian Federal Agency for Science and Innovations and The Russian
Foundation for Basic Research;
Ministry of Education of Slovakia;
Department of Science and Technology, South Africa;
Centro de Investigaciones Energeticas, Medioambientales y Tecnologicas (CIEMAT), E-Infrastructure shared between Europe and Latin America (EELA), Ministerio de Econom\'{i}a y Competitividad (MINECO) of Spain, Xunta de Galicia (Conseller\'{\i}a de Educaci\'{o}n),
Centro de Aplicaciones Tecnológicas y Desarrollo Nuclear (CEA\-DEN), Cubaenerg\'{\i}a, Cuba, and IAEA (International Atomic Energy Agency);
Swedish Research Council (VR) and Knut $\&$ Alice Wallenberg
Foundation (KAW);
Ukraine Ministry of Education and Science;
United Kingdom Science and Technology Facilities Council (STFC);
The United States Department of Energy, the United States National
Science Foundation, the State of Texas, and the State of Ohio;
Ministry of Science, Education and Sports of Croatia and  Unity through Knowledge Fund, Croatia.
Council of Scientific and Industrial Research (CSIR), New Delhi, India
    %%%%%%% done by webmaster team
\end{acknowledgement}

%%%%%%%% Bibliography (In case of using bibtex generate the bbl requested by arXiv)
\bibliographystyle{utphys}   % Put here the style file name for the paper, i.e.apsrev4-1, utphys
\bibliography{citations}
%\input {bibliography.tex}  

%%%%%%%%% appendix with author list
\newpage
\appendix

\section{The ALICE Collaboration}
\label{app:collab}

% Collaboration: CERN-LHC-ALICE
% Generation Date is 2015/Jun/04

% How to use:
%%%%%%%%% appendix with author list
%\appendix
%\section{The ALICE Collaboration}
%\label{app:collab}
%\input{authors-list.tex}  %%%%%%% get the latest version before submitting

\begingroup
\small
\begin{flushleft}
J.~Adam\Irefn{org40}\And
D.~Adamov\'{a}\Irefn{org83}\And
M.M.~Aggarwal\Irefn{org87}\And
G.~Aglieri Rinella\Irefn{org36}\And
M.~Agnello\Irefn{org111}\And
N.~Agrawal\Irefn{org48}\And
Z.~Ahammed\Irefn{org132}\And
S.U.~Ahn\Irefn{org68}\And
I.~Aimo\Irefn{org94}\textsuperscript{,}\Irefn{org111}\And
S.~Aiola\Irefn{org137}
% ...
\renewcommand\labelenumi{\textsuperscript{\theenumi}~}

\section*{Affiliation notes}
\renewcommand\theenumi{\roman{enumi}}
\begin{Authlist}
\item \Adef{0}Deceased
\item \Adef{idp3821664}{Also at: M.V. Lomonosov Moscow State University, D.V. Skobeltsyn Institute of Nuclear, Physics, Moscow, Russia}
\item \Adef{idp5957472}{Also at: University of Kansas, Lawrence, Kansas, United States}
\end{Authlist}

\section*{Collaboration Institutes}
\renewcommand\theenumi{\arabic{enumi}~}
\begin{Authlist}

\item \Idef{org1}A.I. Alikhanyan National Science Laboratory (Yerevan Physics Institute) Foundation, Yerevan, Armenia
\item \Idef{org2}Benem\'{e}rita Universidad Aut\'{o}noma de Puebla, Puebla, Mexico
\item \Idef{org3}Bogolyubov Institute for Theoretical Physics, Kiev, Ukraine
\item \Idef{org4}Bose Institute, Department of Physics and Centre for Astroparticle Physics and Space Science (CAPSS), Kolkata, India
\item \Idef{org5}Budker Institute for Nuclear Physics, Novosibirsk, Russia
\item \Idef{org6}California Polytechnic State University, San Luis Obispo, California, United States
\item \Idef{org7}Central China Normal University, Wuhan, China
\item \Idef{org8}Centre de Calcul de l'IN2P3, Villeurbanne, France
\item \Idef{org9}Centro de Aplicaciones Tecnol\'{o}gicas y Desarrollo Nuclear (CEADEN), Havana, Cuba
\item \Idef{org10}Centro de Investigaciones Energ\'{e}ticas Medioambientales y Tecnol\'{o}gicas (CIEMAT), Madrid, Spain
\item \Idef{org11}Centro de Investigaci\'{o}n y de Estudios Avanzados (CINVESTAV), Mexico City and M\'{e}rida, Mexico
\item \Idef{org12}Centro Fermi - Museo Storico della Fisica e Centro Studi e Ricerche ``Enrico Fermi'', Rome, Italy
\item \Idef{org13}Chicago State University, Chicago, Illinois, USA
\item \Idef{org14}China Institute of Atomic Energy, Beijing, China
\item \Idef{org15}Commissariat \`{a} l'Energie Atomique, IRFU, Saclay, France
\item \Idef{org16}COMSATS Institute of Information Technology (CIIT), Islamabad, Pakistan
\item \Idef{org17}Departamento de F\'{\i}sica de Part\'{\i}culas and IGFAE, Universidad de Santiago de Compostela, Santiago de Compostela, Spain
\item \Idef{org18}Department of Physics and Technology, University of Bergen, Bergen, Norway
\item \Idef{org19}Department of Physics, Aligarh Muslim University, Aligarh, India
\item \Idef{org20}Department of Physics, Ohio State University, Columbus, Ohio, United States
\item \Idef{org21}Department of Physics, Sejong University, Seoul, South Korea
\item \Idef{org22}Department of Physics, University of Oslo, Oslo, Norway
\item \Idef{org23}Dipartimento di Elettrotecnica ed Elettronica del Politecnico, Bari, Italy
\item \Idef{org24}Dipartimento di Fisica dell'Universit\`{a} 'La Sapienza' and Sezione INFN Rome, Italy
\item \Idef{org25}Dipartimento di Fisica dell'Universit\`{a} and Sezione INFN, Cagliari, Italy
\item \Idef{org26}Dipartimento di Fisica dell'Universit\`{a} and Sezione INFN, Trieste, Italy
\item \Idef{org27}Dipartimento di Fisica dell'Universit\`{a} and Sezione INFN, Turin, Italy
\item \Idef{org28}Dipartimento di Fisica e Astronomia dell'Universit\`{a} and Sezione INFN, Bologna, Italy
\item \Idef{org29}Dipartimento di Fisica e Astronomia dell'Universit\`{a} and Sezione INFN, Catania, Italy
\item \Idef{org30}Dipartimento di Fisica e Astronomia dell'Universit\`{a} and Sezione INFN, Padova, Italy
\item \Idef{org31}Dipartimento di Fisica `E.R.~Caianiello' dell'Universit\`{a} and Gruppo Collegato INFN, Salerno, Italy
\item \Idef{org32}Dipartimento di Scienze e Innovazione Tecnologica dell'Universit\`{a} del  Piemonte Orientale and Gruppo Collegato INFN, Alessandria, Italy
\item \Idef{org33}Dipartimento Interateneo di Fisica `M.~Merlin' and Sezione INFN, Bari, Italy
\item \Idef{org34}Division of Experimental High Energy Physics, University of Lund, Lund, Sweden
\item \Idef{org35}Eberhard Karls Universit\"{a}t T\"{u}bingen, T\"{u}bingen, Germany
\item \Idef{org36}European Organization for Nuclear Research (CERN), Geneva, Switzerland
\item \Idef{org37}Excellence Cluster Universe, Technische Universit\"{a}t M\"{u}nchen, Munich, Germany
\item \Idef{org38}Faculty of Engineering, Bergen University College, Bergen, Norway
\item \Idef{org39}Faculty of Mathematics, Physics and Informatics, Comenius University, Bratislava, Slovakia
\item \Idef{org40}Faculty of Nuclear Sciences and Physical Engineering, Czech Technical University in Prague, Prague, Czech Republic
\item \Idef{org41}Faculty of Science, P.J.~\v{S}af\'{a}rik University, Ko\v{s}ice, Slovakia
\item \Idef{org42}Faculty of Technology, Buskerud and Vestfold University College, Vestfold, Norway
\item \Idef{org43}Frankfurt Institute for Advanced Studies, Johann Wolfgang Goethe-Universit\"{a}t Frankfurt, Frankfurt, Germany
\item \Idef{org44}Gangneung-Wonju National University, Gangneung, South Korea
\item \Idef{org45}Gauhati University, Department of Physics, Guwahati, India
\item \Idef{org46}Helsinki Institute of Physics (HIP), Helsinki, Finland
\item \Idef{org47}Hiroshima University, Hiroshima, Japan
\item \Idef{org48}Indian Institute of Technology Bombay (IIT), Mumbai, India
\item \Idef{org49}Indian Institute of Technology Indore, Indore (IITI), India
\item \Idef{org50}Inha University, Incheon, South Korea
\item \Idef{org51}Institut de Physique Nucl\'eaire d'Orsay (IPNO), Universit\'e Paris-Sud, CNRS-IN2P3, Orsay, France
\item \Idef{org52}Institut f\"{u}r Informatik, Johann Wolfgang Goethe-Universit\"{a}t Frankfurt, Frankfurt, Germany
\item \Idef{org53}Institut f\"{u}r Kernphysik, Johann Wolfgang Goethe-Universit\"{a}t Frankfurt, Frankfurt, Germany
\item \Idef{org54}Institut f\"{u}r Kernphysik, Westf\"{a}lische Wilhelms-Universit\"{a}t M\"{u}nster, M\"{u}nster, Germany
\item \Idef{org55}Institut Pluridisciplinaire Hubert Curien (IPHC), Universit\'{e} de Strasbourg, CNRS-IN2P3, Strasbourg, France
\item \Idef{org56}Institute for Nuclear Research, Academy of Sciences, Moscow, Russia
\item \Idef{org57}Institute for Subatomic Physics of Utrecht University, Utrecht, Netherlands
\item \Idef{org58}Institute for Theoretical and Experimental Physics, Moscow, Russia
\item \Idef{org59}Institute of Experimental Physics, Slovak Academy of Sciences, Ko\v{s}ice, Slovakia
\item \Idef{org60}Institute of Physics, Academy of Sciences of the Czech Republic, Prague, Czech Republic
\item \Idef{org61}Institute of Physics, Bhubaneswar, India
\item \Idef{org62}Institute of Space Science (ISS), Bucharest, Romania
\item \Idef{org63}Instituto de Ciencias Nucleares, Universidad Nacional Aut\'{o}noma de M\'{e}xico, Mexico City, Mexico
\item \Idef{org64}Instituto de F\'{\i}sica, Universidad Nacional Aut\'{o}noma de M\'{e}xico, Mexico City, Mexico
\item \Idef{org65}iThemba LABS, National Research Foundation, Somerset West, South Africa
\item \Idef{org66}Joint Institute for Nuclear Research (JINR), Dubna, Russia
\item \Idef{org67}Konkuk University, Seoul, South Korea
\item \Idef{org68}Korea Institute of Science and Technology Information, Daejeon, South Korea
\item \Idef{org69}KTO Karatay University, Konya, Turkey
\item \Idef{org70}Laboratoire de Physique Corpusculaire (LPC), Clermont Universit\'{e}, Universit\'{e} Blaise Pascal, CNRS--IN2P3, Clermont-Ferrand, France
\item \Idef{org71}Laboratoire de Physique Subatomique et de Cosmologie, Universit\'{e} Grenoble-Alpes, CNRS-IN2P3, Grenoble, France
\item \Idef{org72}Laboratori Nazionali di Frascati, INFN, Frascati, Italy
\item \Idef{org73}Laboratori Nazionali di Legnaro, INFN, Legnaro, Italy
\item \Idef{org74}Lawrence Berkeley National Laboratory, Berkeley, California, United States
\item \Idef{org75}Lawrence Livermore National Laboratory, Livermore, California, United States
\item \Idef{org76}Moscow Engineering Physics Institute, Moscow, Russia
\item \Idef{org77}National Centre for Nuclear Studies, Warsaw, Poland
\item \Idef{org78}National Institute for Physics and Nuclear Engineering, Bucharest, Romania
\item \Idef{org79}National Institute of Science Education and Research, Bhubaneswar, India
\item \Idef{org80}Niels Bohr Institute, University of Copenhagen, Copenhagen, Denmark
\item \Idef{org81}Nikhef, Nationaal instituut voor subatomaire fysica, Amsterdam, Netherlands
\item \Idef{org82}Nuclear Physics Group, STFC Daresbury Laboratory, Daresbury, United Kingdom
\item \Idef{org83}Nuclear Physics Institute, Academy of Sciences of the Czech Republic, \v{R}e\v{z} u Prahy, Czech Republic
\item \Idef{org84}Oak Ridge National Laboratory, Oak Ridge, Tennessee, United States
\item \Idef{org85}Petersburg Nuclear Physics Institute, Gatchina, Russia
\item \Idef{org86}Physics Department, Creighton University, Omaha, Nebraska, United States
\item \Idef{org87}Physics Department, Panjab University, Chandigarh, India
\item \Idef{org88}Physics Department, University of Athens, Athens, Greece
\item \Idef{org89}Physics Department, University of Cape Town, Cape Town, South Africa
\item \Idef{org90}Physics Department, University of Jammu, Jammu, India
\item \Idef{org91}Physics Department, University of Rajasthan, Jaipur, India
\item \Idef{org92}Physik Department, Technische Universit\"{a}t M\"{u}nchen, Munich, Germany
\item \Idef{org93}Physikalisches Institut, Ruprecht-Karls-Universit\"{a}t Heidelberg, Heidelberg, Germany
\item \Idef{org94}Politecnico di Torino, Turin, Italy
\item \Idef{org95}Purdue University, West Lafayette, Indiana, United States
\item \Idef{org96}Pusan National University, Pusan, South Korea
\item \Idef{org97}Research Division and ExtreMe Matter Institute EMMI, GSI Helmholtzzentrum f\"ur Schwerionenforschung, Darmstadt, Germany
\item \Idef{org98}Rudjer Bo\v{s}kovi\'{c} Institute, Zagreb, Croatia
\item \Idef{org99}Russian Federal Nuclear Center (VNIIEF), Sarov, Russia
\item \Idef{org100}Russian Research Centre Kurchatov Institute, Moscow, Russia
\item \Idef{org101}Saha Institute of Nuclear Physics, Kolkata, India
\item \Idef{org102}School of Physics and Astronomy, University of Birmingham, Birmingham, United Kingdom
\item \Idef{org103}Secci\'{o}n F\'{\i}sica, Departamento de Ciencias, Pontificia Universidad Cat\'{o}lica del Per\'{u}, Lima, Peru
\item \Idef{org104}Sezione INFN, Bari, Italy
\item \Idef{org105}Sezione INFN, Bologna, Italy
\item \Idef{org106}Sezione INFN, Cagliari, Italy
\item \Idef{org107}Sezione INFN, Catania, Italy
\item \Idef{org108}Sezione INFN, Padova, Italy
\item \Idef{org109}Sezione INFN, Rome, Italy
\item \Idef{org110}Sezione INFN, Trieste, Italy
\item \Idef{org111}Sezione INFN, Turin, Italy
\item \Idef{org112}SSC IHEP of NRC Kurchatov institute, Protvino, Russia
\item \Idef{org113}SUBATECH, Ecole des Mines de Nantes, Universit\'{e} de Nantes, CNRS-IN2P3, Nantes, France
\item \Idef{org114}Suranaree University of Technology, Nakhon Ratchasima, Thailand
\item \Idef{org115}Technical University of Ko\v{s}ice, Ko\v{s}ice, Slovakia
\item \Idef{org116}Technical University of Split FESB, Split, Croatia
\item \Idef{org117}The Henryk Niewodniczanski Institute of Nuclear Physics, Polish Academy of Sciences, Cracow, Poland
\item \Idef{org118}The University of Texas at Austin, Physics Department, Austin, Texas, USA
\item \Idef{org119}Universidad Aut\'{o}noma de Sinaloa, Culiac\'{a}n, Mexico
\item \Idef{org120}Universidade de S\~{a}o Paulo (USP), S\~{a}o Paulo, Brazil
\item \Idef{org121}Universidade Estadual de Campinas (UNICAMP), Campinas, Brazil
\item \Idef{org122}University of Houston, Houston, Texas, United States
\item \Idef{org123}University of Jyv\"{a}skyl\"{a}, Jyv\"{a}skyl\"{a}, Finland
\item \Idef{org124}University of Liverpool, Liverpool, United Kingdom
\item \Idef{org125}University of Tennessee, Knoxville, Tennessee, United States
\item \Idef{org126}University of the Witwatersrand, Johannesburg, South Africa
\item \Idef{org127}University of Tokyo, Tokyo, Japan
\item \Idef{org128}University of Tsukuba, Tsukuba, Japan
\item \Idef{org129}University of Zagreb, Zagreb, Croatia
\item \Idef{org130}Universit\'{e} de Lyon, Universit\'{e} Lyon 1, CNRS/IN2P3, IPN-Lyon, Villeurbanne, France
\item \Idef{org131}V.~Fock Institute for Physics, St. Petersburg State University, St. Petersburg, Russia
\item \Idef{org132}Variable Energy Cyclotron Centre, Kolkata, India
\item \Idef{org133}Vin\v{c}a Institute of Nuclear Sciences, Belgrade, Serbia
\item \Idef{org134}Warsaw University of Technology, Warsaw, Poland
\item \Idef{org135}Wayne State University, Detroit, Michigan, United States
\item \Idef{org136}Wigner Research Centre for Physics, Hungarian Academy of Sciences, Budapest, Hungary
\item \Idef{org137}Yale University, New Haven, Connecticut, United States
\item \Idef{org138}Yonsei University, Seoul, South Korea
\item \Idef{org139}Zentrum f\"{u}r Technologietransfer und Telekommunikation (ZTT), Fachhochschule Worms, Worms, Germany
\end{Authlist}
\endgroup

  %%%%%%% done by webmaster team
\end{document}